\documentclass[preprintnumbers,article,amsmath,amssymb,floatfix,10pt,prd,twocolumn,
superscriptaddress,nofootinbib]{revtex4-2}
\usepackage{bm}
\usepackage{amsfonts}
\usepackage{latexsym}
\usepackage[latin1]{inputenc}
\usepackage{graphicx}
\usepackage{amsmath}
\usepackage{palatino}
\usepackage{mathpazo}
\usepackage{textcomp}
\usepackage{float}
\usepackage{booktabs}
\usepackage{dcolumn}
\usepackage{ragged2e}
\usepackage{hyperref}
\hypersetup{colorlinks,citecolor=blue}
\hypersetup{colorlinks=true,linkcolor=red,filecolor=magenta,    urlcolor=blue}
\usepackage{amsmath}
\usepackage{xcolor}
\usepackage{orcidlink}
\usepackage{epsfig}
\usepackage{caption}
\usepackage{subcaption}
\usepackage{commath}
\def\jnl@style{\it}
\def\aaref@jnl#1{{\jnl@style#1}}

\def\aaref@jnl#1{{\jnl@style#1}}

\def\aj{\aaref@jnl{AJ}}                   % Astronomical Journal
\def\apj{\aaref@jnl{ApJ}}                 % Astrophysical Journal
\def\apjl{\aaref@jnl{ApJ}}                % Astrophysical Journal, Letters
\def\apjs{\aaref@jnl{ApJS}}               % Astrophysical Journal, Supplement
\def\apss{\aaref@jnl{Ap\&SS}}             % Astrophysics and Space Science
\def\aap{\aaref@jnl{A\&A}}                % Astronomy and Astrophysics
\def\aapr{\aaref@jnl{A\&A~Rev.}}          % Astronomy and Astrophysics Reviews
\def\aaps{\aaref@jnl{A\&AS}}              % Astronomy and Astrophysics, Supplement
\def\mnras{\aaref@jnl{Mon.~Not.~Roy.~Astron.~Soc.}}             % Monthly Notices of the RAS
\def\prd{\aaref@jnl{Phys.~Rev.~D}}        % Physical Review D
\def\prc{\aaref@jnl{Phys.~Rev.~C}}  % Physical Review C
\def\prl{\aaref@jnl{Phys.~Rev.~Lett.}}    % Physical Review Letters
\def\qjras{\aaref@jnl{QJRAS}}             % Quarterly Journal of the RAS
\def\skytel{\aaref@jnl{S\&T}}             % Sky and Telescope
\def\ssr{\aaref@jnl{Space~Sci.~Rev.}}     % Space Science Reviews
\def\zap{\aaref@jnl{ZAp}}                 % Zeitschrift fuer Astrophysik
\def\nat{\aaref@jnl{Nature}}              % Nature
\def\aplett{\aaref@jnl{Astrophys.~Lett.}} % Astrophysics Letters
\def\apspr{\aaref@jnl{Astrophys.~Space~Phys.~Res.}} % Astrophysics Space Physics Research
\def\physrep{\aaref@jnl{Phys.~Rep.}}      % Physics Reports
\def\physscr{\aaref@jnl{Phys.~Scr}}       % Physica Scripta
\def\commat{\aaref@jnl{Comm.~Math.~Phys.}}              % Communications in Mathematical Physics
\def\science{\aaref@jnl{Science}}               % Science
\def\cqg{\aaref@jnl{Classical Quant.~Grav.}}            % Classical and Quantum Gravity
\def\jpcs{\aaref@jnl{JPCS}}                                     % Journal of Physics Conference Series
\def\ijmpd{\aaref@jnl{Int.~J.~Mod.~Phys.~D}}                    % International Journal of Modern Physics D
\def\grg{\aaref@jnl{Gen.~Relat.~Gravit.}}               % General Relativity and Gravitation
\def\rpp{\aaref@jnl{Rep.~Prog.~Phys.}}          % Reports on Progress in Physics
\def\npa{\aaref@jnl{Nucl.~Phys.~A}}        % Nuclear Physics A
\def\lrr{\aaref@jnl{Living Rev.~Rel.}}                   % Living reviews in relativity
\def\jcap{\aaref@jnl{J.~Cosmology Astropart.~Phys.}}    % Journal of cosmology and astroparticle physics
\def\rmp{\aaref@jnl{Rev.~Mod.~Phys.}}   %Reviews of modern physics
\def\epjc{\aaref@jnl{Eur.~Phys.~J.~C}} 
\def\plb{\aaref@jnl{~Phy.~Lett.~B}} 
\def\mpla{\aaref@jnl{Mod.~Phy.~Lett.~A}} 
\def\arxiv{\aaref@jnl{arxiv.org}}

\allowdisplaybreaks[1]
\begin{document}
%\color{red}
\color{black}       %% For one column
\title{A model-independent method with phantom divide line crossing in Weyl-type $f(Q,T)$ gravity}
%\end{document}

\author{M. Koussour\orcidlink{0000-0002-4188-0572}}
\email{pr.mouhssine@gmail.com}
\affiliation{Quantum Physics and Magnetism Team, LPMC, Faculty of Science Ben
M'sik,\\
Casablanca Hassan II University,
Morocco.}

%
%%%%%%%%%%%%%%%%%%%%%%%%%%%%%%%%%%%%%  DATE  %%%%%%%%%%%%%%%%%%%%%%%%%%%%%%%%%%%%

\begin{abstract}
We investigate the history of the cosmos using an extension of symmetric
teleparallel gravity, namely Weyl-type $f(Q,T)$ gravity, where $Q$
represents the non-metricity scalar of space-time in the standard Weyl form,
completely specified by the Weyl vector $w_{\mu }$, and $T$ represents the
trace of the matter energy-momentum tensor. We derive the Hubble parameter
from the proposed form of the time-dependent deceleration parameter $q=A-%
\frac{B}{H^{2}}$, where $A$ and $B$ are free constants and then use a
model-independent method to apply it to the Friedmann equations of Weyl-type 
$f(Q,T)$ gravity. Further, we determine the best fit values of the model
parameters using new Hubble sets of data of 31 points and Type Ia Supernovae
(SNe Ia) sets of data of 1048 points. Finally, we examine the behavior of
the effective equation of state~(EoS)~parameter and we observe that the best
fit values of the model parameters supports a crossing of the phantom divide
line i.e. $\omega _{eff}=-1$ from $\omega _{eff}>-1$ (quintessence phase) to 
$\omega _{eff}<-1$ (phantom phase). According to the current study,
Weyl-type $f(Q,T)$ gravity can give an alternative to dark energy (DE) in
solving the existing cosmic acceleration.
\end{abstract}
\date{\today}
\maketitle

%%%%%%%%%%%%%%%%%%%%%%%%%%%%%%%%%%%%%%%%%%%%%%%%%%%%%%%%%%%%%%%%%%%%%%%%
%%%%%%%%%%%%%%%        Introduction        %%%%%%%%%%%%%%%%%%%%%%%%%%%%%
%%%%%%%%%%%%%%%%%%%%%%%%%%%%%%%%%%%%%%%%%%%%%%%%%%%%%%%%%%%%%%%%%%%%%%%%
\section{Introduction}

\label{section 1}

Type Ia Supernovae observations (SNe Ia) indicate an accelerated expansion
of the cosmos/universe \cite{Perlmutter/1999, Riess/1998, Riess/2004}, and
numerous astrophysical observational evidences have shown that the cosmos
undergoes both early inflation and late-time accelerated expansion \cite%
{Spergel/2007, Koivisto/2006, Daniel/2008}. In modern cosmology, this is
thought to be generated by a mysterious kind of energy known as dark energy
(DE), which has a positive energy density and a negative pressure. The
cosmological constant ($\Lambda $) is the most basic candidate for DE. This
comprises the hurdles of fine-tuning and cosmic coincidence \cite{L}.
Further research reveals that various forms of DE exist, such as
quintessence, phantom, tachyon, dilation, and interacting dark energy models
such as holographic and agegraphic models \cite{Ratra/1988, Caldwell/1998,
Caldwell/2002, Caldwell/2003, tachyon, HDE}. Further, the cosmic viscosity
is effective in playing the function of DE candidate, forcing the cosmos to
accelerate \cite{Koussour1, Koussour2}. According to Wilkinson Microwave
Anisotropy Probe (WMAP) observations, DE accounts for 68.3\% of all energy
in the cosmos, dark matter accounts for 26.8\%, and baryonic matter accounts
for the remaining 4.9\% \cite{Planck}. The DE is typically characterized by
an "equation of state (EoS)" parameter $\omega =\frac{p}{\rho }$, which is
the ratio of spatially homogenous DE pressure $p$ to energy density $\rho $.
For cosmic acceleration, $\omega <-\frac{1}{3}$ is necessary. The most basic
explanation for DE is a $\Lambda $ with $\omega _{\Lambda }=-1$, $-1<\omega
<-\frac{1}{3}$ denotes the quintessence model, and $\omega <-1$ is the
phantom era. The present EoS parameter value from observational data such as
SNe Ia data $\omega _{0}=-1.084\pm 0.063$, WMAP data $\omega _{0}=-1.073\pm
_{0.089}^{0.090}$, favors the $\Lambda $CDM scenario. The observable value
of the EoS parameter also favors the quintessence and phantom DE scenarios.

Another approach to describe the current acceleration of the cosmos is to
change the geometry of space-time. We can do this by changing the
Einstein-Hilbert action in General Relativity (GR). One of the most basic
possibilities is to include an arbitrary function of the curvature scalar $R$
in the action, which results in the $f(R)$ gravity \cite{R1, R2, R3}.
Following curvature representations, there are two further approaches:
Teleparallel Gravity (TG), for which gravitational interaction is determined
by the torsion $T$ \cite{T1, T2, T3, T4, T5}. Einstein also utilised
symmetric teleparallel gravity (STG) in an attempt to unify field theories.
It takes into consideration vanishing $R$ and $T$ as well as non-vanishing
non-metricity $Q$. We will concentrate on modified gravity based on
non-metricity, which is a quantity that examines the variation of the length
of a vector in transport parallel process. The STG, also known as the $f(Q)$
gravity, is a lately popular modified gravity \cite{Q1, Q2, Q3, Q4, Q5}.
Furthermore, the $f(Q,T)$ gravity \cite{QT1} is a new extension of $f(Q)$
gravity that is based on the non-minimal coupling between the non-metricity $%
Q$ and the trace of the energy-momentum tensor $T$. The $f(Q,T)$ theory is
built in the same way as the $f(R,T)$ theory. Various investigations have
proven that $f(Q,T)$ gravity is beneficial in describing the current
acceleration of the cosmos and providing a viable solution to the DE
challenge. Arora et al. \cite{QT2} studied if $f(Q,T)$ gravity can solve the
late-time acceleration problem without adding a new sort of DE.
Bhattacharjee et al. \cite{QT3}\ investigated baryogenesis in the context of 
$f(Q,T)$ gravity. The transient behavior of the model with the general form
of $f(Q,T)$ gravity may be determined \cite{QT4}, and the linear case model
parameters may be restricted using 31 Hubble data points and 57 SNe Ia data 
\cite{QT5}.

Similarly, Yixin et al. \cite{Weyl1} investigated Weyl-type $f(Q,T)$ gravity
theory in the context of correct Weyl geometry. However, in the current
approach to $f(Q,T)$ type gravity theories, we have expressed the
non-metricity $Q$ using Weyl geometry prescriptions, where the covariant
divergence of the metric tensor is represented by the product of the metric
tensor and the Weyl vector $w_{\mu }$, i.e. $\nabla _{\lambda }g_{\mu \nu
}=-w_{\lambda }g_{\mu \nu }$. The non-metricity scalar is immediately
connected to the square of the Weyl vector as follows: $Q=-6w^{2}$. As a
result, the Weyl vector and the metric tensor characterize all of the
geometric features of the theory. Yixin et al. \cite{Weyl1} investigated the
cosmological consequences of the Weyl-type $f(Q,T)$ theory for three
distinct classes of models. The obtained solutions represent both
accelerating and decelerating evolutionary periods of the cosmos. It is
observed that Weyl-type $f(Q,T)$ gravity is a plausible contender for
describing the early and late stages of cosmic history. Yang et al. \cite%
{Weyl2} used the Weyl-type $f(Q,T)$ theory to obtain the geodesic and
Raychaudhuri equations. Gadbail et al. \cite{Weyl3} examined the power-law
cosmology\ in the context of Weyl-type $f(Q,T)$ theory. The influence of
viscosity on cosmic evolution has also been investigated in the Weyl type $%
f(Q,T)$ gravity framework \cite{Weyl4}. In the Weyl-type $f(Q,T)$ gravity
theory, the divergence-free parametric form of the deceleration parameter is
examined using a functional form $f(Q,T)$ as $f(Q,T)=\alpha Q+\frac{\beta }{%
6\kappa ^{2}}T$ \cite{Weyl5}.

In this paper, we analyze the time-dependent deceleration parameter, and
examine the Friedmann-Lemaitre-Robertson-Walker (FLRW) cosmos in the
Weyl-type $f(Q,T)$ gravity using a functional form $f(Q,T)=\alpha Q+\frac{%
\gamma }{6\kappa ^{2}}T$, where $\alpha $ and $\gamma $ are free model
parameters. To get exact solutions to the field equations, the proposed
model is based on the time-dependent deceleration parameter as $q=A-\frac{B}{%
H^{2}}$, where $A$ and $B$ are free constants. The present study is
organized as follows: In Sec. \ref{section 2}, we introduce the basic
concepts of Weyl Type $f(Q,T)$ gravity. Sec. \ref{section 3} discusses the
modified Friedmann equations in Weyl Type $f(Q,T)$ gravity and the Hubble
parameter proposed. In Sec. \ref{section 4}, the observational data and the
approach used to constrain the model parameters are discussed. Also, in the
same section we discuss the behavior of some cosmological parameters such as
the deceleration parameter and effective EoS parameter. The behavior of
energy conditions, including the DEC, NEC, and SEC, are presented in Sec. %
\ref{section 5}. Further, In Sec. \ref{section 6} we discuss the behavior of
cosmic jerk parameter. Finally, in the last section \ref{section 7}, we
summarize our findings briefly.

\section{Basic concepts of Weyl-type $f(Q,T)$ gravity}

\label{section 2}

To begin, we consider the gravitational action in the Weyl-type $f(Q,T)$
defined by \cite{Weyl1} 
\begin{multline}
S=\int \left[ \kappa ^{2}f(Q,T)-\frac{1}{4}W_{\alpha \beta }W^{\alpha \beta
}-\frac{1}{2}m^{2}w_{\alpha }w^{\alpha }+\right.   \label{1} \\
\left. \lambda (R+6\nabla _{\mu }w^{\mu }-6w_{\mu }w^{\mu })+\mathcal{L}_{m}%
\right] \sqrt{-g}d^{4}x.
\end{multline}

Here, $W_{\alpha \beta }=\nabla _{\beta }w_{\alpha }-\nabla _{\alpha
}w_{\beta \text{ }}$represents the field strength tensor of the vector
field, $\kappa ^{2}=\frac{1}{16\pi G}$, $m$ is the particle mass associated
to the vector field, and $\mathcal{L}_{m}$ is the matter Lagrangian. In
addition, the second and third terms in the action are the vector field's
ordinary kinetic term and mass term, respectively. Further, $f(Q,T)\equiv f$
can be expressed as an arbitrary function of $Q$ (non-metricity scalar) and
the trace of the matter-energy-momentum tensor $T$.

The scalar non-metricity is crucial to our theory and is obtained by 
\begin{equation}
Q\equiv -g^{\alpha \beta }\left( L_{\nu \beta }^{\mu }L_{\beta \mu }^{\nu
}-L_{\nu \mu }^{\mu }L_{\alpha \beta }^{\nu }\right) ,  \label{2}
\end{equation}%
where, $L_{\alpha \beta }^{\lambda }$ denotes the tensor of deformation, 
\begin{equation}
L_{\alpha \beta }^{\lambda }=-\frac{1}{2}g^{\lambda \gamma }\left( Q_{\alpha
\gamma \beta }+Q_{\beta \gamma \alpha }-Q_{\gamma \alpha \beta }\right) .
\label{3}
\end{equation}

The Levi-Civita connection and the metric tensor $g_{\alpha \beta }$ in
Riemannian geometry can both be compatible, i.e. $\nabla _{\mu }g_{\alpha
\beta }=0$. In Weyl's geometry, however, this appears to be different, 
\begin{equation}
\overline{Q}_{\mu \alpha \beta }\equiv \overline{\nabla }_{\mu }g_{\alpha
\beta }=\partial _{\mu }g_{\alpha \beta }-\overline{\Gamma }_{\mu \alpha
}^{\rho }g_{\rho \beta }-\overline{\Gamma }_{\mu \beta }^{\rho }g_{\rho
\alpha }=2w_{\mu }g_{\alpha \beta },  \label{4}
\end{equation}%
where, $\overline{\Gamma }_{\alpha \beta }^{\lambda }\equiv \Gamma _{\alpha
\beta }^{\lambda }+g_{\alpha \beta }w^{\lambda }-\delta _{\alpha }^{\lambda
}w_{\beta }-\delta _{\beta }^{\lambda }w_{\alpha }$ and $\Gamma _{\alpha
\beta }^{\lambda }$ represents the Christoffel symbol in terms of the metric
tensor $g_{\alpha \beta }$.

From Eqs. \eqref{2}-\eqref{4}, we can derive the following relationship, 
\begin{equation}
Q=-6w^{2}.  \label{5}
\end{equation}

The effective dynamical mass of the vector field is given by $%
m_{eff}^{2}=m^{2}+12\kappa ^{2}f_{Q}+12\lambda $, where $f_{Q}\equiv \frac{%
\partial f(Q,T)}{\partial Q}$. Thus, we can show that the Lagrange
multiplier field creates an effective current for the vector field. Also,
the field equation is derived from the variation principle with regard to
the metric tensor and Weyl vector on Eq. \eqref{1}, 
\begin{multline}
\frac{1}{2}\left( T_{\alpha \beta }+S_{\alpha \beta }\right) -\kappa
^{2}f_{T}\left( T_{\alpha \beta }+\Theta _{\alpha \beta }\right) =-\frac{%
\kappa ^{2}}{2}g_{\alpha \beta }f  \label{7} \\
-6\kappa ^{2}f_{Q}w_{\alpha }w_{\beta }+\lambda \left( R_{\alpha \beta
}-6w_{\alpha }w_{\beta }+3g_{\alpha \beta }\nabla _{\rho }w^{\rho }\right) 
\\
+3g_{\alpha \beta }w^{\rho }\nabla _{\rho }\lambda -6w_{(\alpha }\nabla
_{\beta )}\lambda +g_{\alpha \beta }\square \lambda -\nabla _{\alpha }\nabla
_{\beta }\lambda ,
\end{multline}%
where 
\begin{equation}
T_{\alpha \beta }\equiv -\frac{2}{\sqrt{-g}}\frac{\delta (\sqrt{-g}L_{m})}{%
\delta g^{\alpha \beta }},  \label{8}
\end{equation}%
and%
\begin{equation}
f_{T}\equiv \frac{\partial f(Q,T)}{\partial T},  \label{9}
\end{equation}%
respectively. In addition, the expression for $\Theta _{\alpha \beta }$ is
written as, 
\begin{equation}
\Theta _{\alpha \beta }\equiv g^{\mu \nu }\frac{\delta T_{\mu \nu }}{\delta
g_{\alpha \beta }}=g_{\alpha \beta }L_{m}-2T_{\alpha \beta }-2g^{\mu \nu }%
\frac{\delta ^{2}L_{m}}{\delta g^{\alpha \beta }\delta g^{\mu \nu }}.
\label{10}
\end{equation}

In this research, $S_{\alpha \beta }$ is the re-scaled energy momentum
tensor of the free Proca field 
\begin{equation}
S_{\alpha \beta }=-\frac{1}{4}g_{\alpha \beta }W_{\rho \sigma }W^{\rho
\sigma }+W_{\alpha \rho }W_{\beta }^{\rho }-\frac{1}{2}m^{2}g_{\alpha \beta
}w_{\rho }w^{\rho }+m^{2}w_{\alpha }w_{\beta }.  \label{11}
\end{equation}

\section{Cosmological model and time-dependent deceleration parameter}

\label{section 3}

In this section, we analyze a FLRW cosmos described by the isotropic,
homogeneous, and spatially flat metric, 
\begin{equation}
ds^{2}=-dt^{2}+a^{2}(t)\left[ dx^{2}+dy^{2}+dz^{2}\right] ,  \label{12}
\end{equation}%
where $a(t)$ denotes the scale factor of the cosmos. Due to spatial
symmetry, the vector field is presumed to be of the type $w_{\alpha }=\left[
\psi (t),0,0,0\right] $. As a result, $w^{2}=w_{\alpha }w^{\alpha }=-\psi
^{2}(t)$ with $Q=-6w^{2}=6\psi ^{2}(t)$.

In this case, we adjust the comoving coordinates system $u^{\alpha }=\left(
-1,0,0,0\right) $ as well as $u^{\alpha }\nabla _{\alpha }=\frac{d}{dt}$ and 
$H=\frac{\dot{a}}{a}$. We further suppose that the Lagrangian of the perfect
fluid is $\mathcal{L}_{m}=p$. The associated energy momentum tensor for the
perfect fluid $T_{\mu \nu }$ and $\Theta _{\nu }^{\mu }$ result in $T_{\nu
}^{\mu }=diag\left( -\rho ,p,p,p\right) $ and $\Theta _{\nu }^{\mu }=\delta
_{\nu }^{\mu }p-2T_{\nu }^{\mu }=diag\left( 2\rho +p,-p,-p,-p\right) $,
where $\rho $ and $p$ represent the energy density and pressure of the
cosmos, respectively.

For the cosmological situation, the generalised Proca equation and the flat
space constraint are written as, 
\begin{eqnarray}
\dot{\psi} &=&\dot{H}+2H^{2}+\psi ^{2}-3H\psi ,  \label{17} \\
\dot{\lambda} &=&\left( -\frac{1}{6}m^{2}-2\kappa ^{2}f_{Q}-2\lambda \right)
\psi =-\frac{1}{6}m_{eff}^{2}\psi ,  \label{18} \\
\partial _{i}\lambda &=&0.  \label{19}
\end{eqnarray}

When we apply an FLRW metric to Eq. \eqref{7}, we can get corresponding
modified Friedmann equations as \cite{Weyl1}, 
\begin{align}
\kappa ^{2}f_{T}(\rho +p)& +\frac{1}{2}\rho =\frac{\kappa ^{2}}{2}f-\left(
6\kappa ^{2}f_{Q}+\frac{1}{4}m^{2}\right) \psi ^{2}  \notag \\
& -3\lambda (\psi ^{2}-H^{2})-3\dot{\lambda}(\psi -H),  \label{20}
\end{align}%
\begin{align}
-\frac{1}{2}p& =\frac{\kappa ^{2}}{2}f+\frac{m^{2}\psi ^{2}}{4}+\lambda
(3\psi ^{2}+3H^{2}+2\dot{H})  \notag \\
& +(3\psi +2H)\dot{\lambda}+\ddot{\lambda}.  \label{21}
\end{align}

We eliminate all derivatives of $\lambda $ using Eqs. (\ref{17}) and (\ref%
{18}), and then we take the sum of the two equations above. As a result, we
have a simplified set of cosmic evolution equations, represented by 
\begin{align}
\frac{1}{2}\left( 1+2\kappa ^{2}f_{T}\right) \rho & +\kappa ^{2}f_{T}p=\frac{%
\kappa ^{2}}{2}f+\frac{m^{2}\psi ^{2}}{4}  \notag \\
& +3\lambda \left( H^{2}+\psi ^{2}\right) -\frac{1}{2}m_{eff}^{2}H\psi ,
\label{22}
\end{align}%
\begin{align}
\frac{1}{2}\left( 1+2\kappa ^{2}f_{T}\right) \left( \rho +p\right) & =\frac{%
m_{eff}^{2}}{6}\left( \dot{\psi}+\psi ^{2}-H\psi \right)  \notag \\
& +2\kappa ^{2}\dot{f_{Q}}\psi -2\lambda \dot{H}.  \label{23}
\end{align}%
where dot $(.)$ indicates the time derivative, $f_{Q}$ and $f_{T}$,
respectively, define the derivative with regard to $Q$ and $T$.

By replacing $\dot{\psi}$ as given by Eqs. (\ref{18}) in (\ref{23}), we get%
\begin{eqnarray}
&&\frac{1}{2}\left( 1+2\kappa ^{2}f_{T}\right) \left( \rho +p\right)
=-2\lambda \left( 1-\frac{m_{eff}^{2}}{12\lambda }\right) \dot{H}+  \notag \\
&&\frac{m_{eff}^{2}}{3}\left( H^{2}+\psi ^{2}-2H\psi \right) +2\kappa ^{2}%
\dot{f}_{Q}\psi .  \label{24}
\end{eqnarray}

In this paper, we consider the functional form $f(Q,T)=\alpha Q+\frac{\gamma 
}{6\kappa ^{2}}T$, where $\alpha $ and $\gamma $ are model parameters. The
functional form is influenced by three free parameters, $\alpha $, $\gamma $%
, and $M^{2}=\frac{m^{2}}{\kappa ^{2}}$, where $M$ is the mass of the Weyl
field and $\kappa ^{2}$ denotes the strength of the Weyl geometry-matter
coupling. In this situation, $M=0.95$ is assumed \cite{Weyl1}. It is crucial
to notice that $\gamma =0$ and $\alpha =-1$ correspond to $f(Q,T)=-Q$, which
is an instance of the successful GR. At $T=0$, the theory is simplified to $%
f(Q)=\alpha Q$ gravity, which is identical to GR and satisfies all Solar
System experiments. Furthermore, using the connection relation $\overline{%
\nabla } _{\lambda }g_{\mu \nu }=-w_{\lambda }g_{\mu \nu }$, we may derive $%
\psi (t)=-6H(t)$, where $H(t)=\frac{\dot{a}}{a}$ is the Hubble parameter.

Using Eqs. \eqref{22} and \eqref{23}, we derive the following formulas for
energy density $\rho $ and pressure $p$ as,

\begin{widetext}
\begin{equation}
\rho =\left( \frac{-9\left( 11\gamma +24\right) \left( 24\alpha +25\right) }{%
4\left( \gamma +2\right) \left( \gamma +3\right) }+\frac{29\gamma +72}{2\left(
2\gamma +3\right) \left( \gamma +2\right) }\right) \,H^{2}-\frac{9\gamma \dot{H}%
}{2\left( 2\gamma +3\right) \left( \gamma +3\right) }\,.  \label{25}
\end{equation}%
and%
\begin{equation}
p=-\left( 36\left( \frac{18}{\gamma +3}\left( \alpha +1\right) +\frac{3M^{2}}{%
2\left( \gamma +3\right) }\right) +\frac{18}{2\gamma +3}\right) \,H^{2}-\frac{%
18\left( \gamma +2\right) \dot{H}}{\left( 2\gamma +3\right) \left( \gamma
+3\right) }\,,  \label{26}
\end{equation}

From Eqs. (\ref{25}) and (\ref{26}), the EoS parameter $\omega_{eff} =\frac{p}{%
\rho }$ can be analytically expressed as,%
\begin{equation}
\omega_{eff} =\frac{-\left( 36\left( \frac{18}{\gamma +3}\left( \alpha +1\right) +%
\frac{3M^{2}}{2\left( \gamma +3\right) }\right) +\frac{18}{2\gamma +3}\right)
\,H^{2}-\frac{18\left( \gamma +2\right) \dot{H}}{\left( 2\gamma +3\right)
\left( \gamma +3\right) }}{\left( \frac{-9\left( 11\gamma +24\right) \left(
24\alpha +25\right) }{4\left( \gamma +2\right) \left( \gamma +3\right) }+\frac{%
29\gamma +72}{2\left( 2\gamma +3\right) \left( \gamma +2\right) }\right)
\,H^{2}-\frac{9\gamma \dot{H}}{2\left( 2\gamma +3\right) \left( \gamma
+3\right) }}.  \label{27}
\end{equation}
\end{widetext}

We introduce the deceleration parameter $q$ to characterize the
accelerated/decelerated aspect of the cosmic expansion,%
\begin{equation}
q=-1-\frac{\overset{.}{H}}{H}.  \label{q}
\end{equation}

From Eqs. (\ref{25}) and (\ref{26}), we need to know $H$ in order to solve $%
\rho $ and $p$. A large variety of parametrization methods have been
examined in the literature with the requirement of theoretical consistency
and observational feasibility. This method is called a model-independent way
to explore DE since it takes into account the parametrizations of any
kinematic variable, such as the Hubble parameter, deceleration parameter,
and EoS parameter, and provides the appropriate extra equation. For a review
of this method, see \cite{Pacif1, Pacif2}. As SNe Ia measurements and other
astronomical observations suggest that the cosmos is accelerating, a
time-dependent deceleration parameter is necessary to explain the transition
from deceleration expansion in the past to accelerating expansion in the
present. In line with this idea, different parametrizations have argued that
deceleration parameter is time-dependent in order to examine various
cosmological difficulties \cite{P1, P2, P3, P4, P5, P6, P7, P8, P9, P10,
P11, P12}. Motivated by the preceding debate, we use a generalization form
of the deceleration parameter proposed in Eq. (\ref{q}) as a function of the
Hubble parameter: $q=A-\frac{B}{H^{2}}$. Here, $A$ is a dimensionless
constant and $B$ has the dimensions of $H^{2}$ \cite{Tiwari}.

Using the above relation for $q$ with Eq. (\ref{q}) to solve the scale
factor and the Hubble parameter, one obtains:%
\begin{equation}
a(t)=\left\{ \sinh \left[ \sqrt{\left( 1+A\right) B}t+c\right] \right\} ^{%
\frac{1}{1+A}},  \label{at}
\end{equation}%
\begin{equation}
H(t)=\sqrt{\frac{B}{1+A}}\coth \left[ \sqrt{\left( 1+A\right) B}t+c\right] ,
\label{Ht}
\end{equation}%
where $c$ is the constant of integration. Also, to establish a direct
comparison between theoretical predictions of the cosmological model and
astronomical observations, we employ the redshift $z$, which is related to $%
a\left( t\right) $ by the relationship $a\left( t\right) =\left( 1+z\right)
^{-1}$. Using Eqs. (\ref{at}) and (\ref{Ht}), we get%
\begin{equation}
H\left( z\right) =\sqrt{\frac{B}{1+A}\left[ \left( 1+z\right) ^{2\left(
1+A\right) }+1\right] .}
\end{equation}

To reduce the number of parameters, we find the Hubble parameter in terms of 
$H_{0}=H(z=0)$,%
\begin{equation}
H\left( z\right) =H_{0}\sqrt{\frac{1}{2}\left[ \left( 1+z\right) ^{2\left(
1+A\right) }+1\right] },  \label{Hz}
\end{equation}%
with $H_{0}$ being the current Hubble parameter value.

The derivative of the Hubble parameter with regard to cosmic time can be
expressed as,%
\begin{equation}
\overset{.}{H}=-\left( 1+z\right) H\left( z\right) \frac{dH}{dz}.
\end{equation}

Inserting Eq. (\ref{Hz}) into Eq. (\ref{q}), we get $q\left( z\right) $ as,%
\begin{equation}
q\left( z\right) =-1+\frac{\left( 1+A\right) \left( 1+z\right) ^{2\left(
A+1\right) }}{1+\left( 1+z\right) ^{2\left( A+1\right) }}.  \label{qz}
\end{equation}

The behavior and fundamental cosmological features of the model given in Eq.
(\ref{Hz}) are completely dependent on the model parameters ($H_{0}$, $A$).
In the next part, we constrain the model parameters ($H_{0}$, $A$) by using
current observational sources of data to investigating the behavior of
cosmological parameters.

\section{Observational data analysis}

\label{section 4}

Now, we assess the feasibility of the model using current observational
data, especially observational Hubble set of data \cite{Sharov} and SNe Ia 
\cite{Scolnic/2018}. We utilize the Pantheon sample for SNe Ia set of data,
which contains 1048 points from the Low-z, SDSS, Pan-STARSS1 (PS1) Medium
Deep Survey, SNLS, and HST surveys \cite{Chang/2019}.

\subsection{Hubble set of data}

First, we use a typical compilation of 57 Hubble data observations acquired
using the differential age approach (DA), BAO and various redshift ranges $%
0.07<z<2.42$ \cite{Sharov}. This approach may be used to calculate the rate
of expansion of the cosmos at redshift $z$. $H(z)=-\frac{dz/dt}{(1+z)}$ may
be computed here. We do this analysis by minimizing 
\begin{equation}
\chi _{Hubble}^{2}=\sum_{i=1}^{31}\frac{\left[ H(z_{i},\Theta
)-H_{obs}(z_{i})\right] ^{2}}{\sigma (z_{i})^{2}},
\end{equation}%
where $H(z_{i},\Theta )$ is the theoretical value for a particular model at
redshifts $z_{i}$, and $\Theta =\left( H_{0},A\right) $ denotes the
parameter space. The observational value and error are represented by $%
H_{obs}(z_{i})$ and $\sigma (z_{i})^{2}$, respectively. Fig. \ref{F_Hz} also
shows an error bar plot of 31 points of $H(z)$ and a comparison of the
considered model with the well-motivated $\Lambda $CDM model. We have
assumed $\Omega _{\Lambda _{0}}=0.7$, $\Omega _{m_{0}}=0.3$, and $H_{0}=69$
km/s/Mpc.

\subsection{SNe Ia set of data}

The Pantheon sample is a composite of many supernova surveys that found SNe
Ia at both high and low redshift. As a result, the entire sample extends
throughout the redshift range $0.01<z<2.26$. The normalized observational
SNe Ia distance modulus is obtained by: $\mu =m_{B}-M_{B}+\alpha x_{1}-\beta
c+\Delta _{M}+\Delta _{B}$, assuming the Tripp estimator \cite{Tripp/1998}
and the light curve fitter, where $m_{B}$, $M_{B}$, and $c$ denote the
measured peak magnitude (at the maximum of the B-band), absolute magnitude,
and $SNe\,$ color, respectively. Furthermore, $\alpha $ and $\beta $ denote
the relationship between luminosity stretch and luminosity color,
respectively. $\Delta _{M}$ and $\Delta _{B}$ also provide distance
adjustments on host galaxy mass and simulation-based anticipated biases.

Scolnic et al. \cite{Scolnic/2018,Kessler/2017} employed the BEAMS with Bias
Corrections (BBC) approach to adjust for errors owing to intrinsic scatter
and selection effects based on realistic SNe Ia simulations. As a result,
the distance modulus is reduced to $\mu =m_{B}-M_{B}$. The following
expressions were utilized in our analysis: 
\begin{eqnarray}
\mu ^{th} &=&5log_{10}\left( \frac{d_{L}(z)}{1Mpc}\right) +25, \\
d_{L}(z) &=&c(1+z)\int_{0}^{z}\frac{dz^{\prime }}{H(z^{\prime },\Theta )}, \\
\chi _{SNeIa}^{2} &=&min\sum_{i,j=1}^{1048}\Delta \mu
_{i}(C_{SNeIa}^{-1})_{ij}\Delta \mu _{j}.
\end{eqnarray}

In Fig. \ref{F_Mu}, we employed error bars to depict 1048 pantheon sample
points and compared our model to the well-accepted $\Lambda $CDM model. We
took $\Omega _{\Lambda _{0}}=0.7$, $\Omega _{m_{0}}=0.3$, and $H_{0}=69$
km/s/Mpc into account.

\subsection{Results}

Further, we employ the total likelihood function to obtain combined
constraints from the Hubble and SNe Ia for the parameters $H_{0}$ and $A$.
The corresponding likelihood and Chi-square functions are now defined by%
\begin{eqnarray}
\mathcal{L}_{Total} &=&\mathcal{L}_{Hubble}\times \mathcal{L}_{SNeIa}, \\
\chi _{Total}^{2} &=&\chi _{Hubble}^{2}+\chi _{SNeIa}^{2}.
\end{eqnarray}

The model parameter constraints are calculated by minimizing the respective
2 using Markov Chain Monte Carlo (MCMC) and the emcee library. Tab. \ref%
{tab1} displays the results. Moreover, the best fit values of $H_{0}$ and $A$
are calculated using Hubble and SNe Ia sets of data, as shown in triangle
plot \ref{F_H} and \ref{F_tot} with $1-\sigma $ and $2-\sigma $ confidence
level. The likelihoods are perfectly suited to Gaussian distributions.

\begin{widetext}

\begin{figure}[H]
\centering
\includegraphics[scale=0.6]{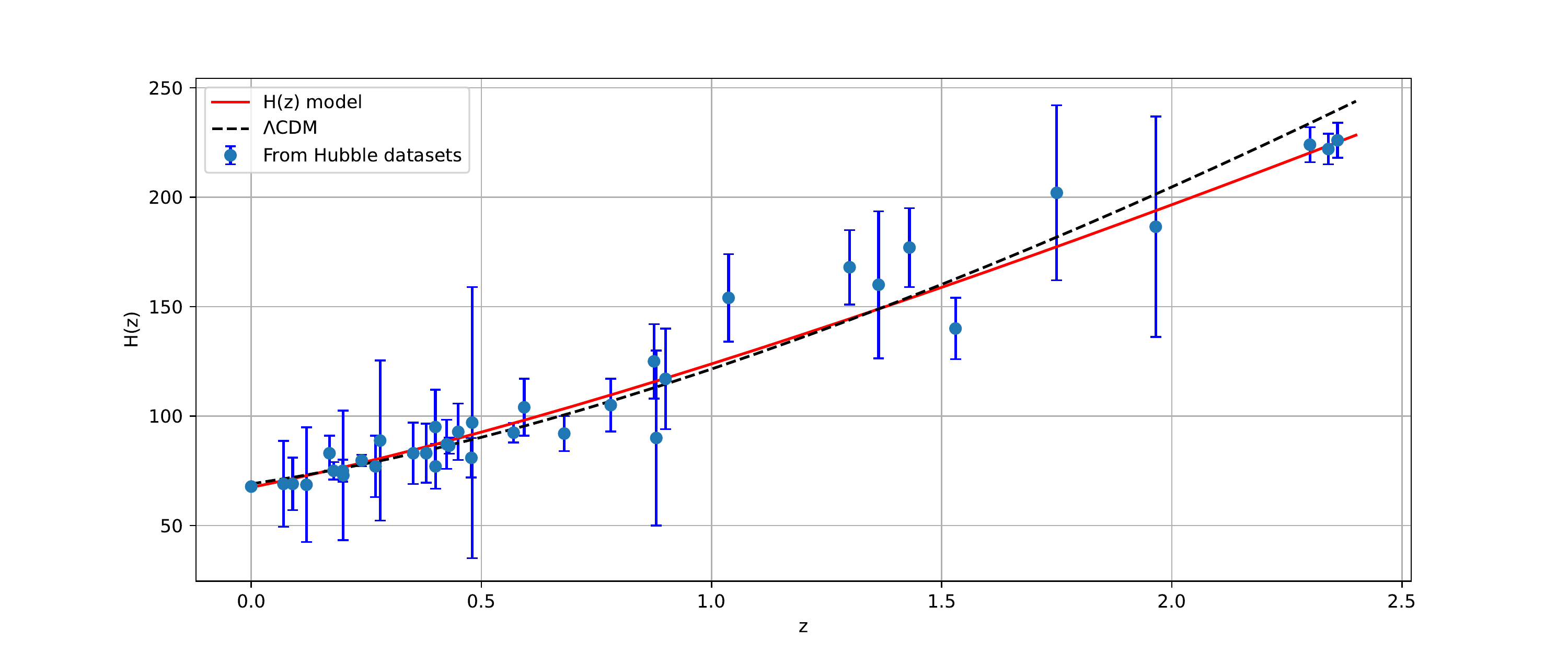}
\caption{Error bar plot of $H$ vs $z$ for our model (red curve) and the $\Lambda$CDM model (black dotted curve). The blue dots represent the 31 Hubble data points.}
\label{F_Hz}
\end{figure}

\begin{figure}[H]
\centering
\includegraphics[scale=0.6]{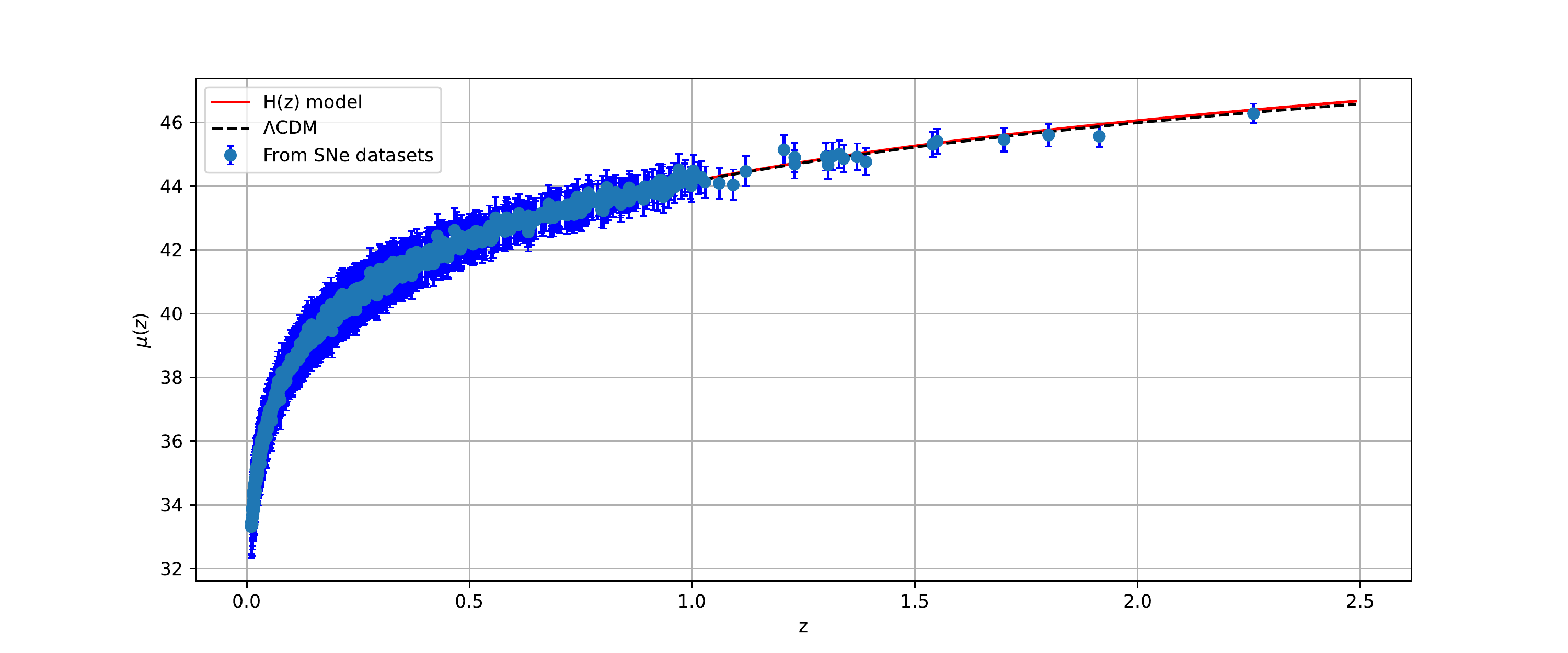}
\caption{Error bar plot of $\mu$ vs $z$ for our model (red curve) and the $\Lambda$CDM model (black dotted curve). The blue dots represent the 1048 pantheon points.}
\label{F_Mu}
\end{figure}

\begin{table}[h!]
\begin{center}
  \caption{Best-fit model parameter values derived from observational sources of data.}
    \label{table1}

\begin{tabular}{|l|c|c|c|c|}
\hline 
Sets of data              & $H_0$ & $A$ & $q_{0}$ & $\omega_{0}$\\
\hline
Hubble             & $67.4^{+1.5}_{-1.5}$ & $0.262^{+0.039}_{-0.039}$ & $-0.369^{+0.0195}_{-0.0195}$ & $-1.01445^{+0.01}_{-0.01}$ \\ 
\hline
SNe Ia           & $67.2^{+2.1}_{-2.2}$ & $0.19^{+0.19}_{-0.18}$ & $-0.405^{+0.095}_{-0.090}$ & $-1.0332^{+0.01}_{-0.01}$ \\
\hline
\end{tabular}
\label{tab1}
\end{center}
\end{table}
\end{widetext}

\begin{widetext}

\begin{figure}[H]
\centering
\includegraphics[scale=0.99]{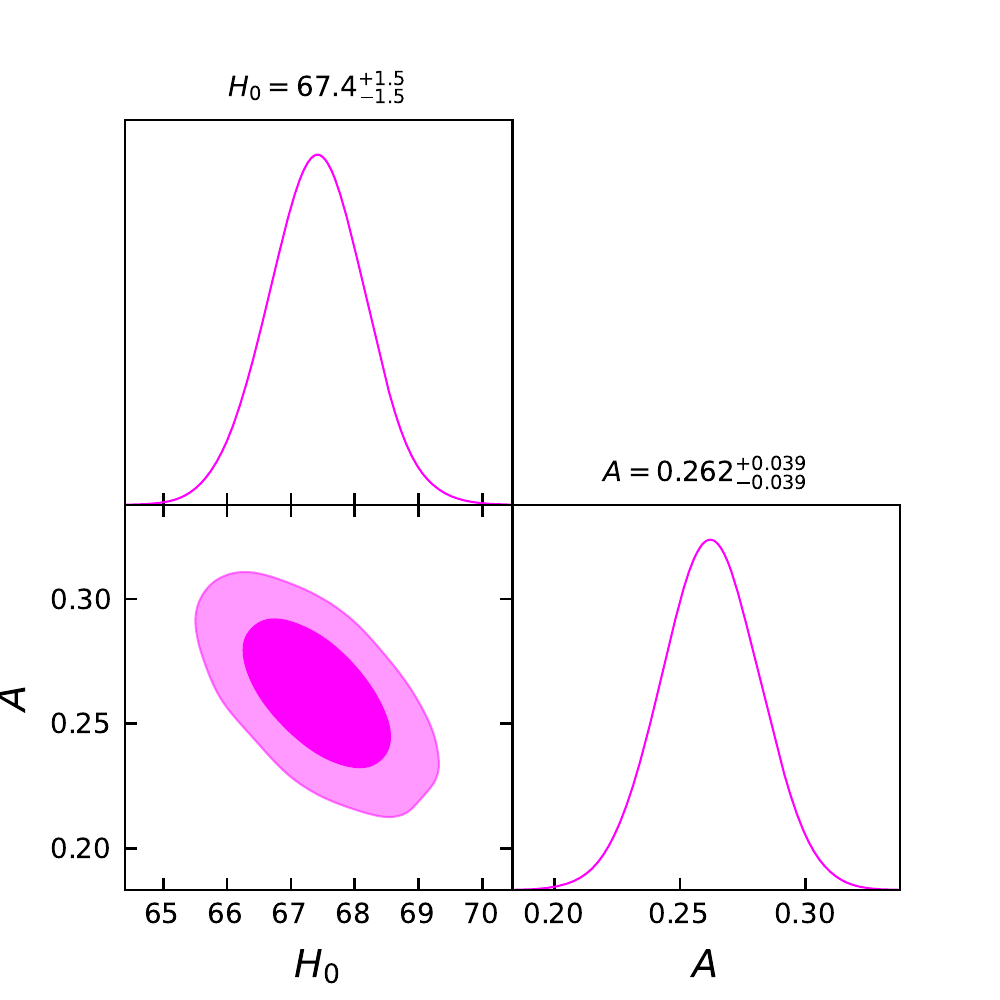}
\caption{The confidence areas of $1-\sigma$ and $2-\sigma$ for the parameters corresponding to the Hubble sets of data.}
\label{F_H}
\end{figure}

\begin{figure}[H]
\centering
\includegraphics[scale=0.99]{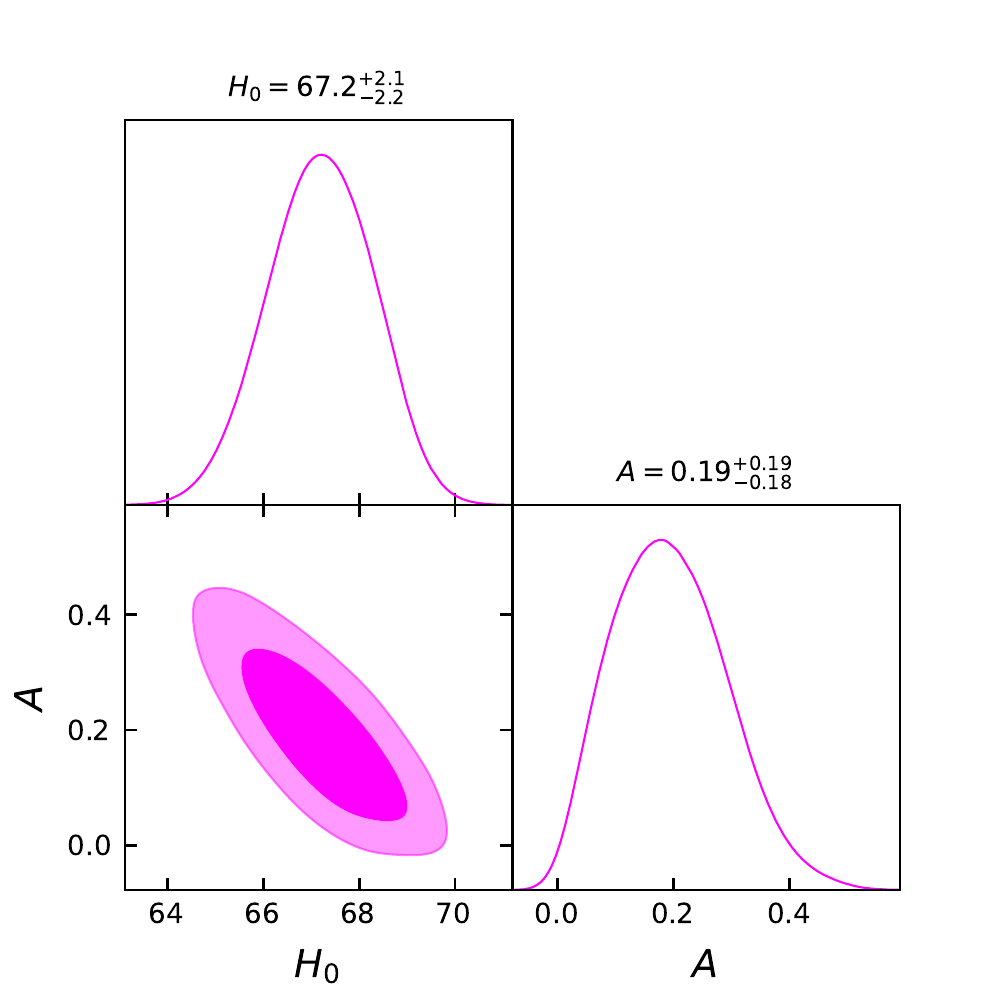}
\caption{The confidence areas of $1-\sigma$ and $2-\sigma$ for the parameters corresponding to the SNe Ia sets of data}
\label{F_tot}
\end{figure}

\end{widetext}

\begin{figure}[tbp]
\centering{\includegraphics[scale=0.6]{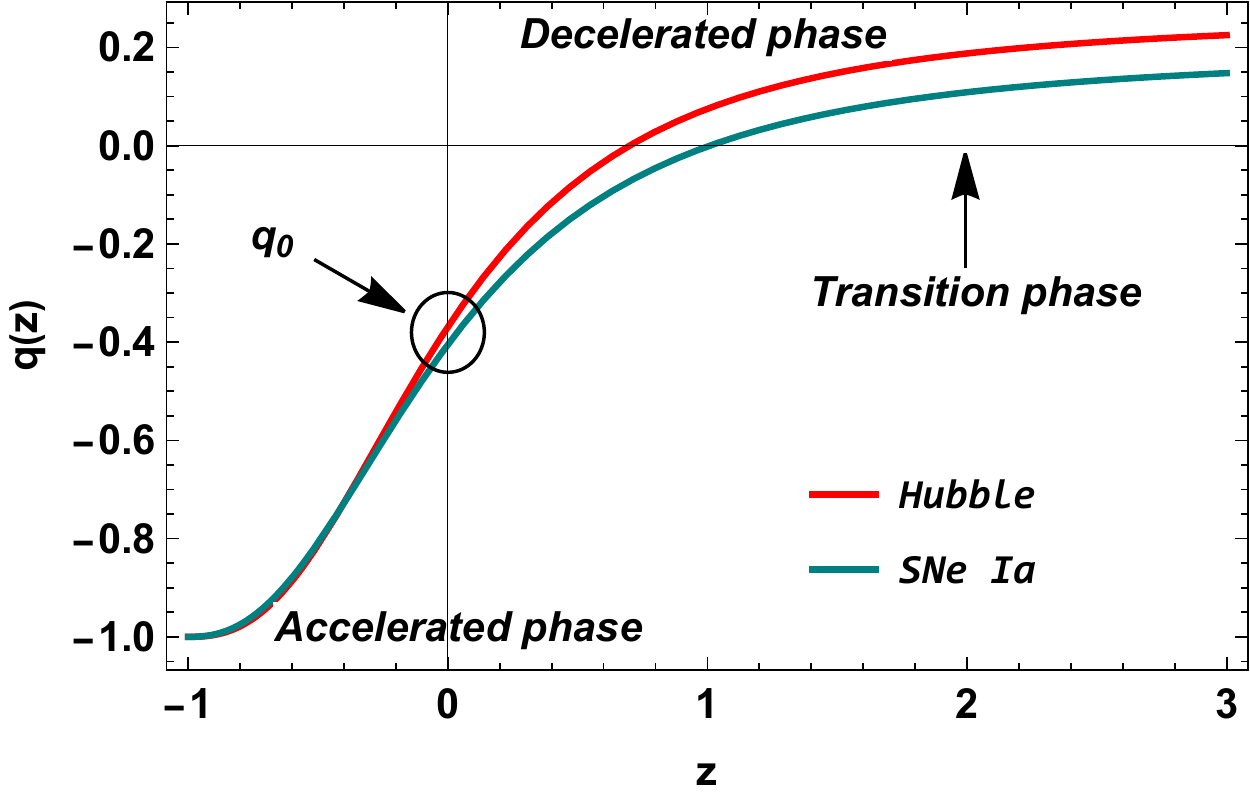}}
\caption{Curve of the deceleration parameter for the proposed model,
according to the values of the parameters constrained by the Hubble, and SNe
Ia sets of data.}
\label{F_qz}
\end{figure}

The observed cosmic acceleration is a new concern, according to cosmological
measurements. In the absence of DE, or if its influence is negligible, the
same model should decelerate in the early phase of the matter era to allow
the structure formation of the cosmos. Thus, to explain the full
evolutionary history of the cosmos, a cosmological model must include also a
decelerated and an accelerated period of expansion. Hence, it is important
to investigate the behavior of the deceleration parameter given in Eq. (\ref%
{qz}). Fig. \ref{F_qz} depicts the evolution of the deceleration parameter
for the corresponding values of model parameters constrained by the Hubble
and SNe Ia sets of data. The deceleration parameter rapidly decreases and
approaches $-1$ asymptotically, indicating de-Sitter-like expansion at late
time ($z\rightarrow -1$). For this scenario, the deceleration parameter
represents a transition from a decelerating expansion phase (i.e.
matter-dominated universe) to the current accelerating phase of the cosmos.
According to Fig. \ref{F_qz}, the value of deceleration parameter is
positive at the beginning of the cosmos ($z>0$) and becomes negative at the
end ($z\rightarrow -1$). The negative value of deceleration parameter
denotes the accelerating expansion of the cosmos. According to recent Planck
data observations, the value of deceleration parameter is in the range $%
-1<q<0$. As a result, our developed model is adequate for describing
the~late time cosmos~history.

\begin{figure}[tbp]
\centering{\includegraphics[scale=0.6]{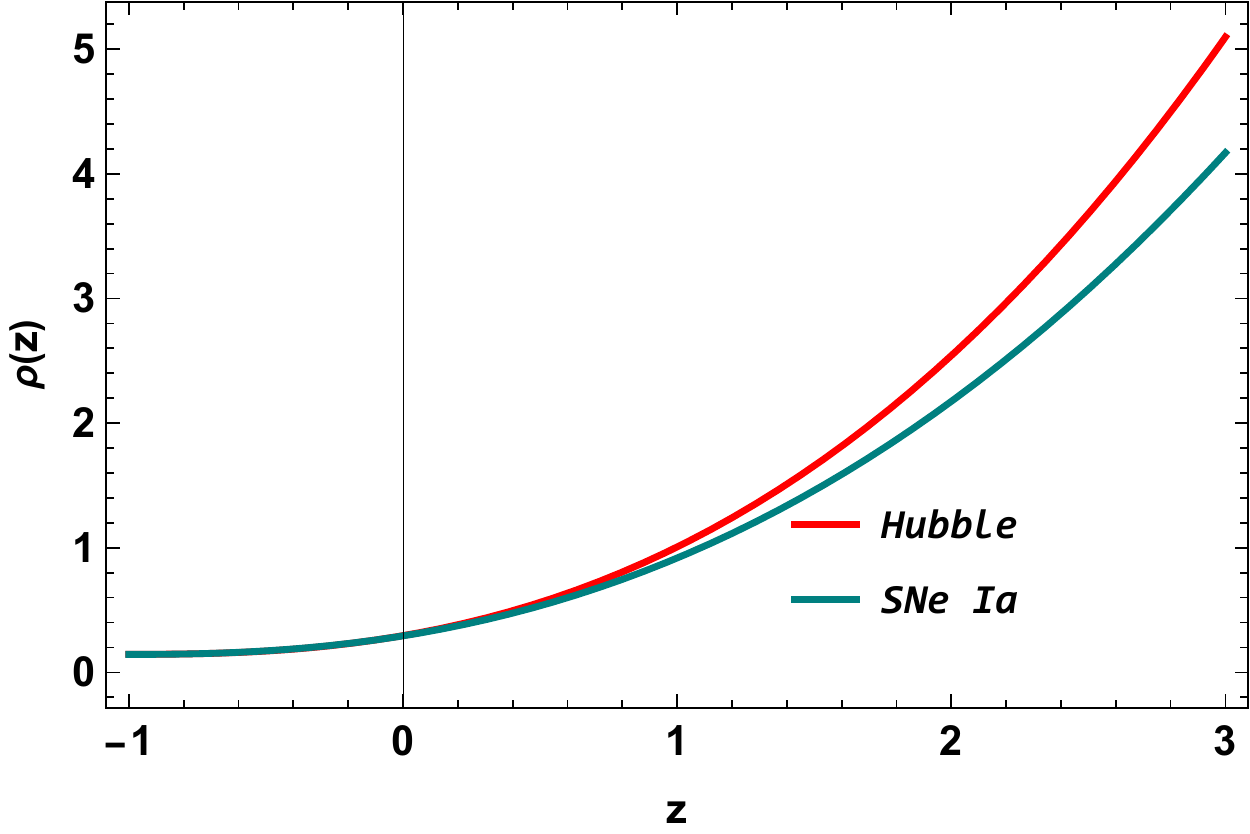}}
\caption{Curve of the energy density for the proposed model, according to
the values of the parameters constrained by the Hubble, and SNe Ia sets of
data.}
\label{F_rho}
\end{figure}

\begin{figure}[tbp]
\centering{\includegraphics[scale=0.6]{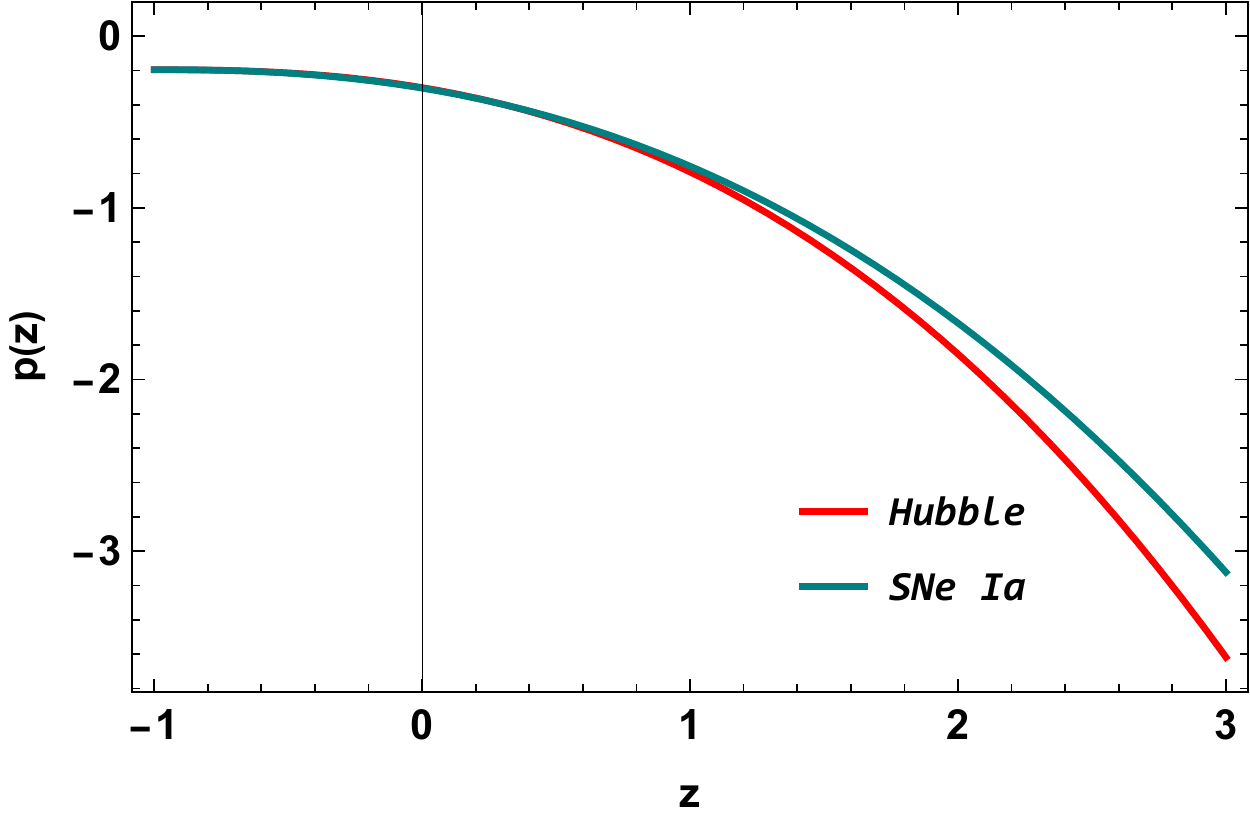}}
\caption{Curve of the pressure for the proposed model, according to the
values of the parameters constrained by the Hubble, and SNe Ia sets of data.}
\label{F_p}
\end{figure}

Because the model parameters $\alpha $ and $\gamma $ are not explicitly
present in the calculation of the Hubble parameter (see Eq. (\ref{Hz})), we
try to fix them to investigate the evolution of the cosmological parameters.
So we used the values $\alpha =-1.053$ and $\gamma =0.5$. Eq. (\ref{25})
expresses the energy density. As seen in Fig. \ref{F_rho}, the energy
density decreases with the history of the cosmos and eventually vanishes.
This shows the expansion of the cosmos. Also, the pressure is expressed by
Eq. (\ref{26}), and its behavior is seen in Fig. \ref{F_p}. According to
Fig. \ref{F_p}, the pressure p remains negative for both two values of the
constrained parameters throughout the history, as predicted, indicating
accelerated expansion of the cosmos.

\begin{figure}[tbp]
\centering{\includegraphics[scale=0.6]{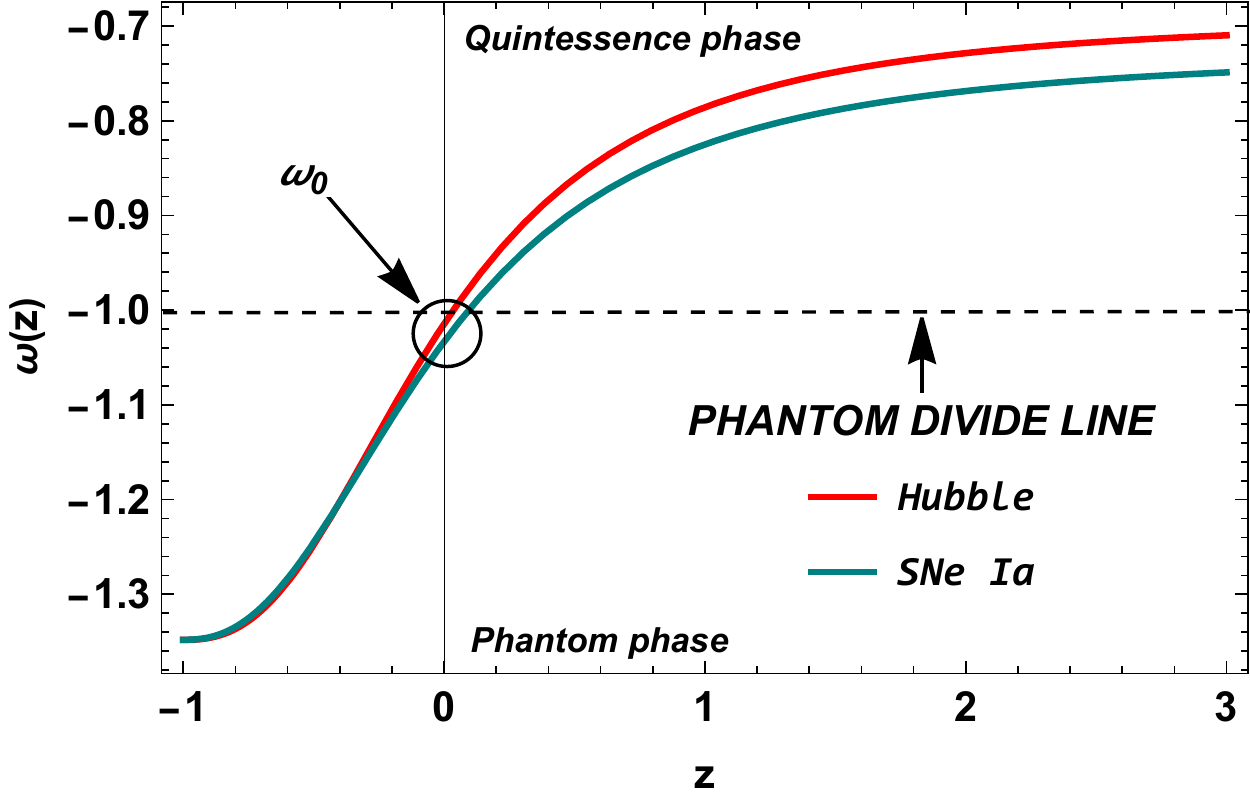}}
\caption{Curve of the effective EoS parameter for the proposed model,
according to the values of the parameters constrained by the Hubble, and SNe
Ia sets of data.}
\label{F_EoS}
\end{figure}

The EoS parameter can be defined as the isotropic pressure $p$ to energy
density $\rho $ ratio, i.e. $\omega =\frac{p}{\rho }$. The EoS of DE can
describe cosmic inflation and accelerated expansion of the cosmos. $\omega <-%
\frac{1}{3}$ is the condition for an accelerating cosmos. In the most basic
example, $\omega =-1$ is according to the cosmological constant, $\Lambda $%
CDM. Furthermore, the values of EoS $\omega =\frac{1}{3}$ and $\omega =0$
indicate a radiation-dominated cosmos and a matter-dominated cosmos,
respectively. If, $-1<\omega <-\frac{1}{3}$, it represents quintessence
model and $\omega <-1$ shows phantom behavior of the model. Fig. \ref{F_EoS}
depicts the behavior of the effective EoS parameter for the parameters
constrained by the Hubble and SNe Ia sets of data. It is clear that the best
fit value of the model parameters supports a crossing of the phantom divide
line i.e. $\omega _{eff}=-1$ from $\omega _{eff}>-1$ (quintessence phase) to 
$\omega _{eff}<-1$ (phantom phase). Also, the current values of the EoS
parameter are presented in Tab. \ref{tab1}. These values clearly show that
the present cosmos is an accelerating phase and lies in the~phantom phase.
Thus, these results are consistent with several cosmological models that
have been proposed in the literature \cite{phantom1, phantom2, phantom3}. 

\section{Energy conditions}

\label{section 5}

The physical parameters which include the deceleration parameter and the EoS
parameter are important in the study of the cosmos. Now, another major
research in current cosmology is on energy conditions (ECs) derived from
Raychaudhuri's equation \cite{ECs1}. ECs provide an important advantage in
the GR for gaining a wide knowledge of the space-time singularity theorem.
Capozziello et al. \cite{ECs2} described in great detail the generalized ECs
in extended theories of gravity. They presented ECs in this article by
contracting timelike and null vectors with regard to the Ricci, Einstein,
and energy-momentum tensors. Capozziello and Laurentis \cite{ECs3}, in
addition to Capozziello et al. \cite{ECs4}, presented a very comprehensive
description of the ECs in modified gravity theories. The main objective of
these ECs is to examine the expansion of the cosmos. There are several types
of ECs, namely null energy condition (NEC), weak energy condition (WEC),
dominant energy condition (DEC), and strong energy condition (SEC). These
ECs are given in Weyl-type $f(Q,T)$ modified theory of gravity with defined
energy density $\rho $ and pressure $p$ as follows: (a) NEC: $\rho +p\geq 0$%
. The violation of the NEC results in the violation of the second law of
thermodynamics. The NEC is violated if $\omega $ is in phantom region $%
\omega <-1$. (b) WEC: $\rho \geq 0,\rho +p\geq 0$. This condition for energy
density decreases. (c) DEC: $\rho \geq 0,\rho \pm p\geq 0$. (d) SEC: $\rho
+p\geq 0,\rho +3p\geq 0$. The violation of NEC leads in the violation of
residual ECs, which symbolizes the decrease of energy density with the
expansion of the cosmos. Further, the violation of SEC indicates the
acceleration of the cosmos.

\begin{figure}[tbp]
\centering{\includegraphics[scale=0.6]{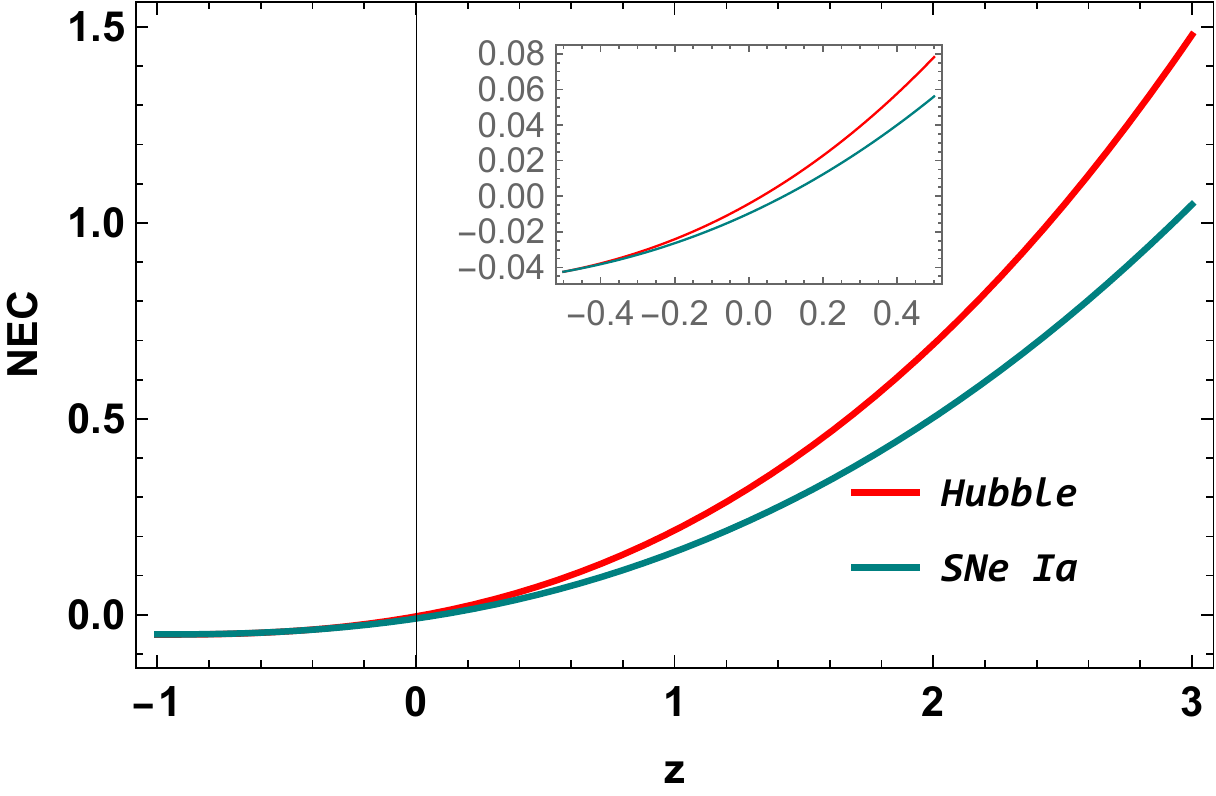}}
\caption{Curve of the NEC for the proposed model, according to the values of
the parameters constrained by the Hubble, and SNe Ia sets of data.}
\label{F_NEC}
\end{figure}

\begin{figure}[tbp]
\centering{\includegraphics[scale=0.6]{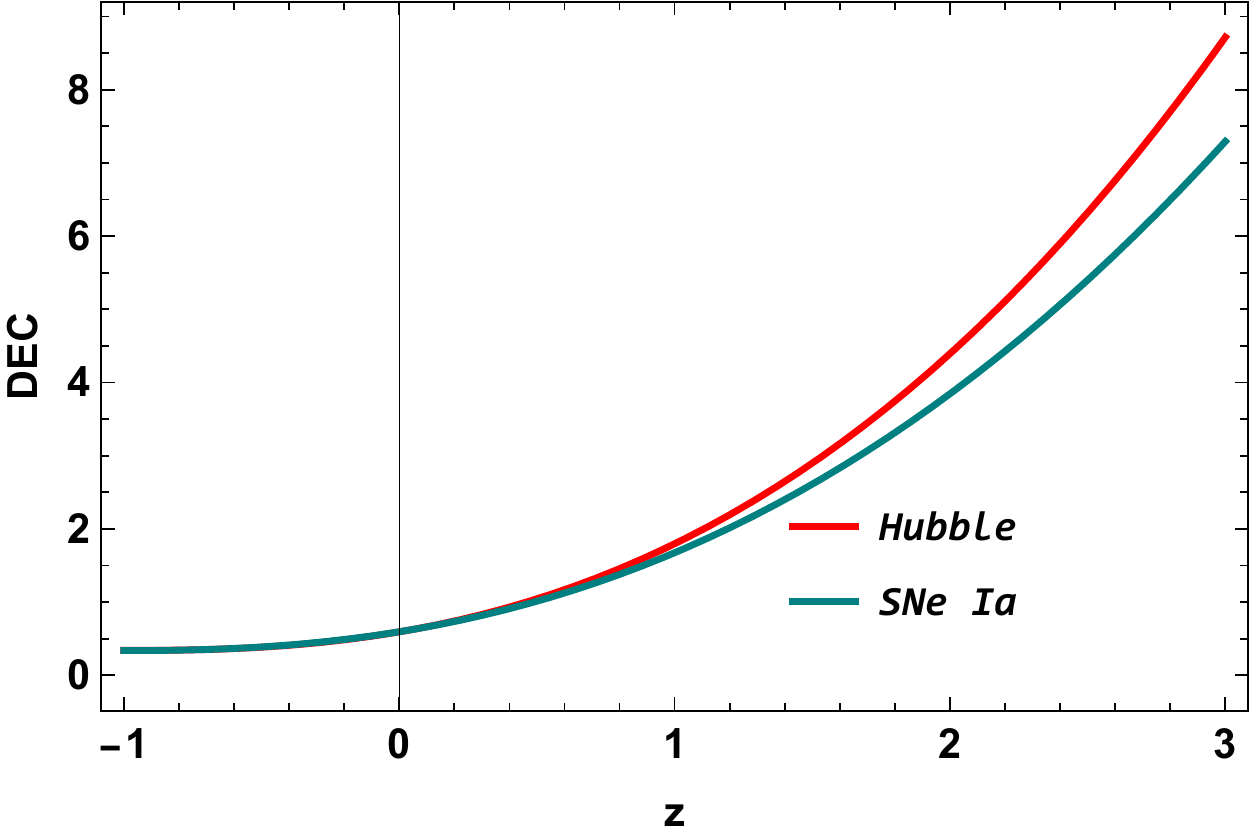}}
\caption{Curve of the DEC for the proposed model, according to the values of
the parameters constrained by the Hubble, and SNe Ia sets of data.}
\label{F_DEC}
\end{figure}

\begin{figure}[tbp]
\centering{\includegraphics[scale=0.6]{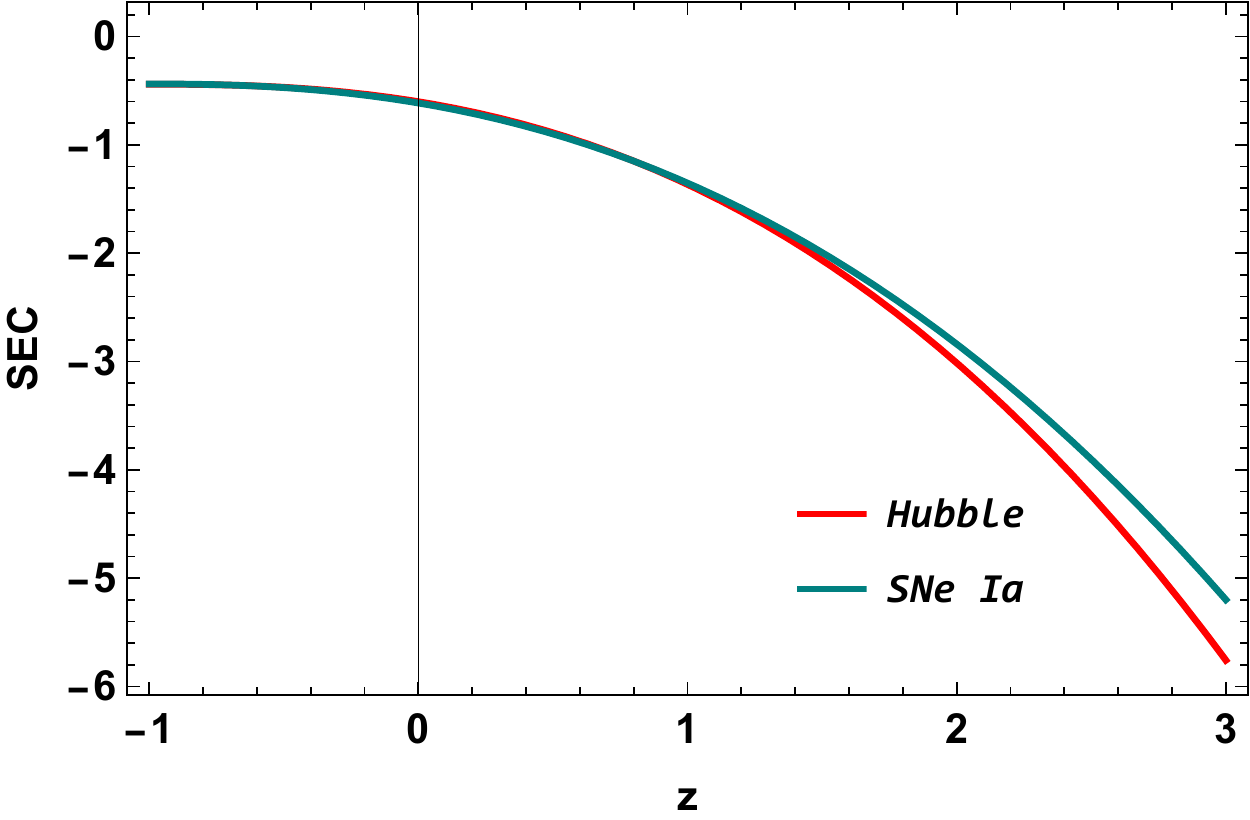}}
\caption{Curve of the SEC for the proposed model, according to the values of
the parameters constrained by the Hubble, and SNe Ia sets of data.}
\label{F_SEC}
\end{figure}

Figs. \ref{F_NEC}, \ref{F_DEC}, and \ref{F_SEC} show graphs of ECs with
respect to redshift for Weyl-type $f(Q,T)$ gravity, and we can also see that 
$\rho -p\geq 0$, which indicates the DEC is satisfied. In addition, Figs. %
\ref{F_NEC} and \ref{F_SEC} show that $\rho +p\leq 0$ and $\rho +3p\leq 0$
at $z=0$ (present), leads in a violation of the NEC and SEC, respectively.
As a result, a violation of the SEC causes the cosmos to accelerate and
behaves like phantom model.

\section{Cosmic jerk parameter}

\label{section 6}

The jerk parameter is regarded as a crucial quantity for understanding the
dynamics of the cosmos. The cosmic jerk parameter can define models that are
near to $\Lambda $CDM. In addition, the transition from the decelerating to
the accelerating phase of the cosmos is thought to be caused by a cosmic
jerk. This cosmos transition happens for several models with a positive jerk
parameter and a negative deceleration parameter. The value of jerk for the
flat $\Lambda $CDM model is $j=1$ \cite{jerk}.

The cosmic jerk parameter is a dimensionless quantity that contains the
third order derivative of the scale factor with regard to cosmic time and is
written as,%
\begin{equation}
j=\frac{\dddot{a}}{aH^{3}}=2q^{2}+q-\frac{\dot{q}}{H}.  \label{j}
\end{equation}

Using Eqs. (\ref{Hz}) and (\ref{qz}) in Eq. (\ref{j}) we get the expression
for the cosmic jerk parameter as,%
\begin{equation}
j\left( z\right) =\frac{2A^{2}(z+1)^{2A+2}+A(z+1)^{2A+2}+1}{%
z^{2}(z+1)^{2A}+2z(z+1)^{2A}+(z+1)^{2A}+1}.
\end{equation}

\begin{figure}[tbp]
\centering{\includegraphics[scale=0.6]{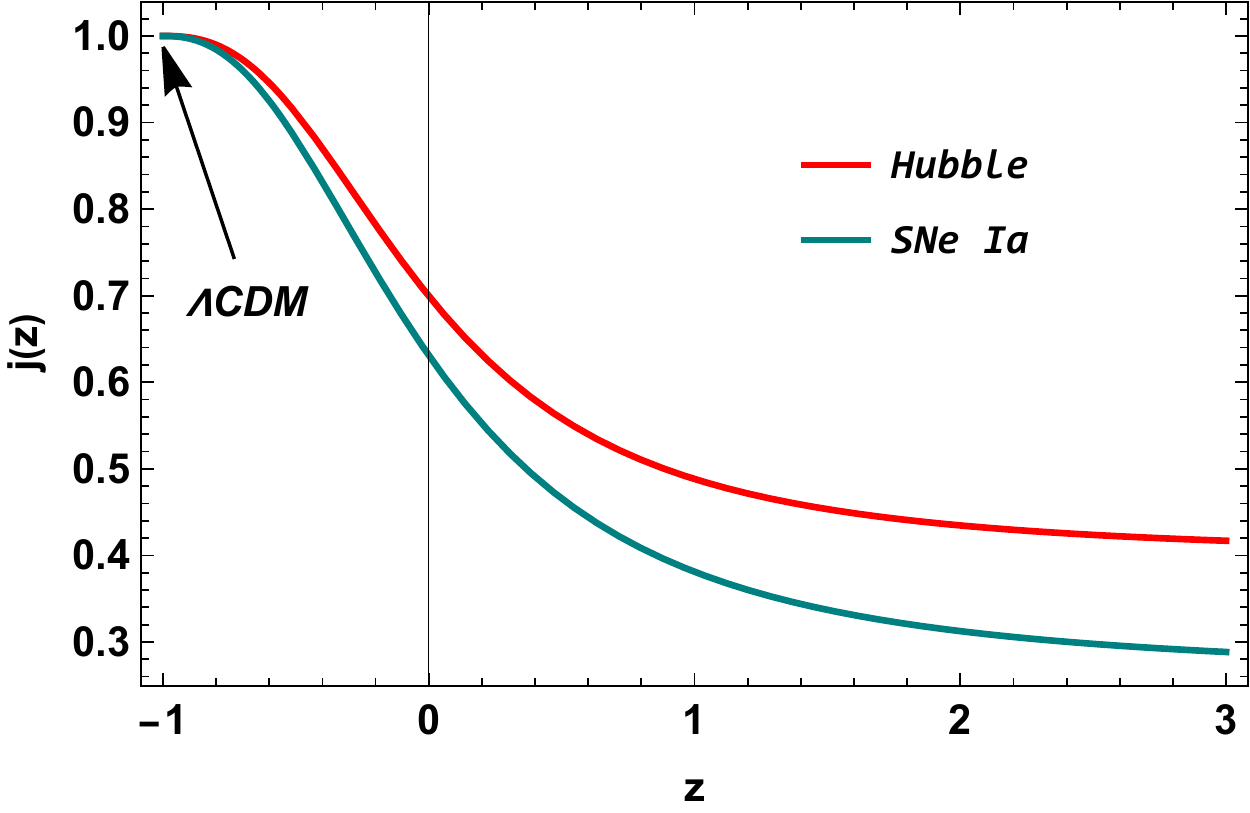}}
\caption{Curve of the cosmic jerk parameter for the proposed model,
according to the values of the parameters constrained by the Hubble, and SNe
Ia sets of data.}
\label{F_j}
\end{figure}

In Fig. \ref{F_j}, we exhibited the cosmic jerk parameter for various values
of the model parameters constrained by the Hubble, and SNe Ia sets of data.
One can see that the cosmic jerk parameter remains positive throughout the
cosmos and is equivalent to the $\Lambda $CDM model at $z\rightarrow -1$ for
the Hubble and SNe Ia sets of data. It is important to note that our model
is similar to the $\Lambda $CDM model.

\section{Conclusions}

\label{section 7}

We proposed an accelerated cosmic model that depicts phantom behavior in the
current and late stages of evolution. To begin, we derived modified
Friedmann equations with an assumed form of the function $f(Q,T)$,
especially $f(Q,T)=\alpha Q+\frac{\gamma }{6\kappa ^{2}}T$, where $Q$
represents the non-metricity scalar of space-time in the standard Weyl form,
completely specified by the Weyl vector $w_{\mu }$, and $T$ represents the
trace of the matter energy-momentum tensor, and to comprehend cosmic
evolution, we used a time-dependent deceleration parameter as $q=A-\frac{B}{%
H^{2}}$, where $A$ and $B$ are free constants. Using the previous equation
for $q$, we can calculate the Hubble parameter in terms of redshift $z$.
Further, we used the MCMC approach to constrain the parameters $A$ and $B$
using the Hubble, and SNe Ia sets of data. Tab. \ref{table1} shows the
best-fit values for the $A$ and $B$ parameters. Then, we examined the
evolution of several cosmological parameters associated with these best-fit
model parameter values. The deceleration parameter in this scenario
indicates a transition from a decelerating expansion phase to the current
accelerating phase of the cosmos. The energy density $\rho $\ of the cosmos
decreases throughout time and finally vanishes. This depicts the expanding
of the cosmos. Furthermore, as expected, the pressure $p$ remains negative
for both two values of the constrained parameters throughout the history,
indicating accelerated expansion of the cosmos. The effective EoS parameter
crosses the phantom divide line i.e. $\omega _{eff}=-1$ from $\omega
_{eff}>-1$ (quintessence phase) to $\omega _{eff}<-1$ (phantom phase).
Furthermore, we observed that the NEC and SEC are violated, while the DEC is
satisfied in Figs. \ref{F_NEC}, \ref{F_DEC}, and \ref{F_SEC}. Violation of
the SEC and NEC causes the cosmos to accelerate and behaves like phantom
model. Finally, to compare our model with the most widely accepted model in
cosmology, we analyzed the cosmological behavior of the cosmic jerk
parameter, which we observed that it remains positive throughout the cosmos
and is equivalent to the $\Lambda $CDM model in the future. According to the
current research, Weyl-type $f(Q,T)$ gravity with a time-dependent
deceleration parameter can provide an alternative to dark energy in solving
the current cosmic acceleration problem.

\textbf{Data availability} There are no new data associated with this
article.

\textbf{Declaration of competing interest} The authors declare that they
have no known competing financial interests or personal relationships that
could have appeared to influence the work reported in this paper.\newline

\section*{acknowledgments}

I am very much grateful to the honorable referee and to the editor for the illuminating suggestions that have significantly improved my work in terms of research quality, and presentation. Also, as I write this, my students at \textbf{Zineb Nefzaouia High School (TCSF-1, TCSF-2, and TCSF-3)} have surprised me with many special gifts on the occasion of obtaining my Ph.D., so I would like to thank them for their support. Particularly noteworthy are two students, \textbf{Fadil Marwa} and \textbf{Makhlouq Achraf}, whom I believe will go on to become excellent physicists. I wish you all the best in your academic careers, my dear students. Thanks for everything.

\end{document}